%% file: sample-authordraft.tex
\documentclass[sigconf]{acmart}

\AtBeginDocument{%
  \providecommand\BibTeX{{%
    \normalfont B\kern-0.5em{\scshape i\kern-0.25em b}\kern-0.8em\TeX}}}

\copyrightyear{2020}
\acmYear{2020}
\setcopyright{acmcopyright}

\acmConference[SIGIR '20] {43rd International ACM SIGIR Conference on Research and Development in Information Retrieval}{July 25--30, 2020}{Virtual Event, China}
\acmBooktitle{43rd International ACM SIGIR Conference on Research and Development in Information Retrieval (SIGIR '20), July 25--30, 2020, Virtual Event, China}
\acmPrice{15.00}
\acmDOI{10.1145/3397271.3401281}
\acmISBN{978-1-4503-8016-4/20/07}

\usepackage{titlesec}
\usepackage{caption}
\usepackage{subfigure}
\usepackage{booktabs}
\usepackage{multirow}
\usepackage{multicol}

\titleformat{\paragraph}[runin]{\bfseries\itshape}{\theparagraph}{1em}{}
\titlespacing*{\paragraph}{0pt}{3.25ex plus 1ex minus .2ex}{1em}

\newcommand{\bb}{\textbf}
\newcommand{\rating}{\mathbf{r}}
\newcommand{\argmin}[1]{\underset{#1}{\operatorname{arg}\,\operatorname{min}}\;}

\makeatletter \renewcommand\paragraph{\@startsection{paragraph}{4}{\z@} {2mm \@plus1ex \@minus.2ex} {-1em} {\normalfont\normalsize\bfseries}} \makeatother

\begin{document}
\fancyhead{}

\title{How Useful are Reviews for Recommendation? A Critical Review and Potential Improvements}

\author{Noveen Sachdeva}
\affiliation{%
  \institution{International Institute of Information Technology}
  \city{Hyderabad, India}
}
\email{ernoveen@gmail.com}
\authornote{Work done during internship at UC San Diego.}

\author{Julian McAuley}
\affiliation{%
  \institution{University of California, San Diego}
  \city{La Jolla, CA, USA}
}
\email{jmcauley@ucsd.edu}

\begin{abstract}
We investigate a growing body of work that seeks to improve recommender systems through the use of review text. Generally, these papers argue that since reviews `explain' users' opinions, they ought to be useful to infer the underlying dimensions that predict ratings or purchases. Schemes to incorporate reviews range from simple regularizers to neural network approaches. Our initial findings reveal several discrepancies in reported results, partly due to (e.g.)~copying results across papers despite changes in experimental settings or data pre-processing. First, we attempt a comprehensive analysis to resolve these ambiguities. Further investigation calls for discussion on a much larger problem about the ``importance" of user reviews for recommendation. Through a wide range of experiments, we observe several cases where state-of-the-art methods fail to outperform existing baselines, especially as we deviate from a few narrowly-defined settings where reviews are useful. We conclude by providing hypotheses for our observations, that seek to characterize under what conditions reviews are likely to be helpful. Through this work, we aim to evaluate the direction in which the field is progressing and encourage robust empirical evaluation.
\end{abstract}

\maketitle

\section{Introduction}
One of the main directions in recommender systems research seeks to improve prediction through the use of `side information,' especially in cold-starting settings where interaction data may be sparse or noisy. A promising and popular direction seeks to make use of \emph{user reviews}, which often exist alongside rating or purchase data. Reviews are a natural source of data to exploit, as
(1) a single review is much more expressive than a single rating; and (2) reviews are specifically intended to `explain' the underlying dimensions behind users' decisions, which ought to be particularly informative in cold-start settings. This setup has been thoroughly explored, generally in the context of rating prediction, where it has been argued that review data can substantially improve recommendation accuracy.

Previously, there largely have been two schools of thought regarding employing user reviews for better recommendation. The first type considers reviews as ``explanations" for the user giving that specific rating and tries to incorporate them into matrix factorization (MF). HFT \cite{hft} is such a model which tries to regularize the latent features being learned through MF by reusing the same latent features for modeling the reviews' likelihood using LDA \cite{lda}. The other type of methods are based on the philosophy that textual reviews are much more expressive than a single rating, and can be used to learn better latent features to perform better MF. \cite{deepconn,narre,transnet,mpcn} are all popular methods which, in some different way, try to extract features from user reviews and item reviews through deep learning architectures like TextCNN \cite{textcnn}, and use these extracted features to perform MF. All the methods used for analysis in this paper are discussed in more detail in Section \ref{all_methods_used}.

\textbf{In this paper,} we argue that the benefit of using reviews for recommendation is overstated, and in particular, the substantial reported gains are only possible under a narrow set of conditions. Our experiments reveal that (1) in most practical cases, recent systems fail to outperform simple baselines (differing from what is usually reported); and (2) many such systems exhibit only a minor change in performance when reviews are masked from the model. Ultimately we conclude that (1) reviews \emph{can} be important, but the recent modeling techniques for reviews are questionable; (2) reviews seem to be more effective when used as a regularizer, rather than as data to extract better latent features; and (3) the community should focus on more consistent empirical evaluation, especially concerning dataset choices, and pre-processing strategies.

Our work also connects to recent discussions \cite{making_progress} on the reproducibility of recent neural methods for recommendation. Note that the topic of this paper is different from \cite{making_progress} since, in addition to the correctness of recent works, we also deal with a more general meta-question about the utility of reviews for recommendation.

\section{Preliminaries}
\subsection{Problem Setup}
Many traditional recommender systems involve learning from a sparse $|\mathcal{U}| \times |\mathcal{I}|$ boolean interaction matrix, constructed using ($u, i$) interaction tuples. For review-based recommendation, it is assumed that with each tuple we also have a numerical rating \footnote{Reviews are rarely used in one-class/implicit feedback settings, partly due to the lack of review data associated with negative instances \cite{mengting_goodreads}.}, $\rating_u^i$ and a textual review $\delta_u^i$: a sequence of tokens (words), where the user $u$ explains their reason for giving the item $i$ the rating value of $\rating_u^i$. 
\vspace{-0.2cm}
\subsection{Methods compared} \label{all_methods_used}

To evaluate the utility of reviews for recommendation, in this work, we consider a wide variety of representative methods from different categories of recommender systems. These range from traditional MF methods, to simpler review-based methods, and finally four ``state-of-the-art'' deep-learning and review-based methods. We omit a discussion on other (not compared) methods due to space considerations. We now list and briefly discuss the methods we use in our experiments (in chronological order of date of publishing):

\paragraph{Bias only:} A naive baseline that assumes the user and item to be independent (i.e.,~considers no mutual interactions). Formally, we learn scalar user and item biases, $\beta_u$ and $\beta_i$ for each user and item, and a global bias $\alpha$. The rating is modeled as: $\hat{\rating}_u^i = \alpha + \beta_u + \beta_i$.

\paragraph{Matrix Factorization (MF) \cite{mf}:} MF  tries to improve upon the bias only model by learning latent features $\gamma_u, \gamma_i$ for users and items respectively. Ratings are modeled as:
$\hat{\rating}_u^i = \alpha + \beta_u + \beta_i + \left( \gamma_u \cdot \gamma_i \right)$.

\paragraph{Hidden Factors and Topics (HFT) \cite{hft}:} Is an initial model that attempts to exploit reviews for better rating prediction. HFT follows a traditional MF setup, with an additional regularizer that models the corpus likelihood using LDA \cite{lda}. We call the regularization function ``$\mathit{lik}$" for notational convenience. Formally:
\vspace{-0.15cm}
\begin{align*}
    \argmin{\alpha, \beta, \gamma} & \sum_{(u, i) ~ \in ~ \mathcal{D}} \left[\hat{\rating}_u^i  - \rating_u^i \right]^2 - \mu \cdot \mathit{lik}(\delta_u^i \in \mathcal{D}~|~\gamma_u, \gamma_i)
\end{align*}

\paragraph{Deep Co-operative Neural Network (DeepCoNN) \cite{deepconn}:}  Was one of the first neural networks proposed for modeling reviews for recommendation. It assumes all reviews given by/to a single user/item to be independent and forms a user/item review document by concatenating all reviews given by/to the user/item. A famous CNN-based architecture---TextCNN \cite{textcnn} is used to extract latent features from the review documents, and finally, the rating is modeled as the output of a neural network conditioned on the extracted latent features. We also consider a version called DeepCoNN++ (not in the original paper) where we learn the global, user, and item biases ($\alpha, \beta_u, \beta_i$) and add it to the neural network's output.

\paragraph{Neural Matrix Factorization (NeuMF) \cite{neural_mf}:} Improves upon traditional MF by modeling the interaction of $\gamma_u, \gamma_i$ with a neural network, $F$. Formally, $\hat{\rating}_u^i = \alpha + \beta_u + \beta_i + F\left(\gamma_u, \gamma_i \right)$. We treat this as a strong non-review-based baseline.

\paragraph{TransNets \cite{transnet}:} In addition to using the user $u$ and item $i$'s review document for extracting latent features, Transnets also uses the current review ($\delta_u^i$) for regularization. Principally, it has two sub-models, one focusing on the sentiment in the given review ($\delta_u^i$) and the other being the same as DeepCoNN. Regularization is performed by minimizing the distance between the latent spaces in both components. We also consider a version---TransNets++ where MF latent features are concatenated to the latent textual features.

\paragraph{Neural Attentive Rating Regression (NARRE) \cite{narre}:}  Primarily improves over DeepCoNN's assumption 
of review independence
by learning an attention weight over individual reviews in the review document. NARRE also uses 
TextCNN to extract features for each review
and learns the global, user, and item biases by default.

\paragraph{Multi-Pointer Co-Attention Networks for Recommendation (MPCN) \cite{mpcn}:} Introduces a deep architecture following the same intuition as NARRE that not every review is equally important, and tries to infer this importance dynamically. 
Unlike NARRE's attention mechanism, MPCN proposes a review-by-review pointer-based learning scheme to infer review importance.

\section{Research Methodology}

\subsection{Datasets} \label{all_data_used}
We use 
(1) six categories from the Amazon review datasets
\footnote{\href{https://cseweb.ucsd.edu/~jmcauley/datasets.html}{https://cseweb.ucsd.edu/~jmcauley/datasets.html}} 
\cite{dataset_paper}, and (2) the BeerAdvocate dataset \cite{beer_dataset}
for running our experiments. These datasets are intended to demonstrate varying levels of sparsity (which we find to be related to the effectiveness of review-based recommendation), with the Amazon datasets generally being the sparsest and the BeerAdvocate being the densest. The data consists of numerous $(u, i, \rating_u^i, \delta_u^i)$ tuples (see statistics in Table~\ref{all_results}) on which we follow a randomized 80:10:10 train/test/validation split. We use the validation set to search for optimal hyper-parameters and report the test set performance of the best performing model.

\paragraph{User/item Pruning:} 
It is typical for existing papers to use $k$-core versions of the datasets, i.e., each node in the bipartite user-item interaction graph has a degree of at least $k$. Doing so saves experimentation time,
and the number of reviews left are significantly reduced (Table~\ref{all_results}). However, doing so---either deliberately or accidentally---favors methods that work well on dense datasets (or poorly on sparse ones). As such, this preprocessing scheme seems to stand against the initial motivations of using review text for recommendation---to perform better for colder users/items. Hence, to assess this inconsistency, we consider both scenarios when we use the 0-core dataset,
and the corresponding $k=5$-core subset.

\paragraph{Textual Preprocessing:} Following the setting in NARRE \cite{narre}, we don't remove stopwords and maintain a vocabulary of the 50K most frequent words.
For performance, we use 64-dimensional word2vec embeddings trained using Gensim\footnote{\href{https://radimrehurek.com/gensim/}{https://radimrehurek.com/gensim/}}.
Following the original implementations of the respective methods, for DeepCoNN, we cap/pad the user/item document length to be 1000 tokens and for other methods, we cap/pad the length of each review to be equal to the top-2 percentile, and fix the number of reviews similarly. 
Note that all test \& validation set reviews were removed while training.

\input{all_results}

\vspace{-0.15cm}
\subsection{Implementation}
We were able to find public implementations of some of the models online and reused them for our experiments. We also implemented \textbf{all} the models ourselves\footnote{\href{https://github.com/noveens/reviews4rec}{Code available at https://github.com/noveens/reviews4rec}}. In case of a mismatch, the better result among the public implementation and ours is reported.

\paragraph{Hyper-parameter search:}
To ensure that inconsistent results are not due to poor hyper-parameter tuning, we conduct thorough hyper-parameter search for all methods on the validation set. The latent dimension was searched in $[1, 4, 8, 25, 50]$, L2 regularization in $[10^{-4}, 10^{-5}, 10^{-6}, 10^{-7}]$, and dropout in $[0.2, 0.4, 0.6, 0.8]$.

\subsection{Experiments}

\paragraph{How do different methods perform on different datasets?} MSE values are reported in Table \ref{all_results}. Surprisingly: (1) the bias-only method performs quite well when compared to MF on the 0-core versions of the Amazon datasets; (2) simple models (like HFT) outperform more sophisticated neural methods like DeepCoNN, NARRE, MPCN, etc. on \emph{all} of the 0-core datasets and are comparable on the 5-core subsets; and (3) the ``++" versions of DeepCoNN \& Transnets have large differences in MSE compared to their simpler counterparts, owing to the strong performance of the bias only method.

\begin{figure}[ht]
    \centering
    \vspace{-0.15cm}
    \includegraphics[width=8.3cm, height=3.2cm]{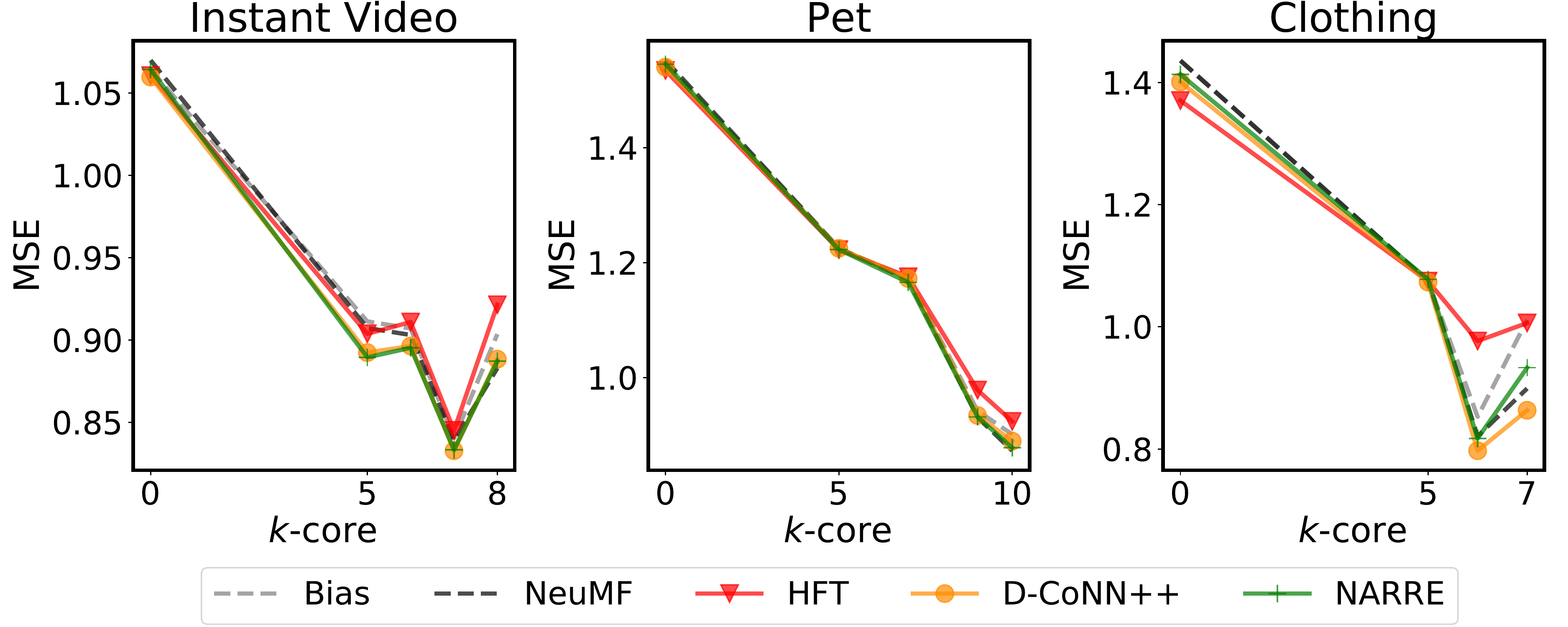}
    \vspace{-0.15cm}
    \caption{Performance comparison varying $k$.}
    \label{density}
    \vspace{-0.35cm}
\end{figure}

\paragraph{How does for performance change with varying sparsity?} To better understand performance change of different methods with varying sparsity, instead of just zero and 5-core subsets, we evaluate the changes in MSE on an even more extensive range of $k$-core subsets. We increase $k$ until we have no users/items left. Because of space limitations and plot clarity, only a few representative results are reported in Figure~\ref{density}. As expected, most methods perform better with increasing density. Comparing to other methods, HFT becomes \emph{comparatively} worse as we increase the density as we have more reviews for each user and item, such that it is logical to model reviews as features rather than regularizers. This argument is also supported by the \emph{relative} increase in the performance of text-as-features-based methods like DeepCoNN and NARRE. We also note the increase in the \emph{relative} performance of MF methods with the bias-only methods as the density increases. 


\begin{figure}[ht]
    \centering
    \includegraphics[width=8.3cm, height=5.7cm]{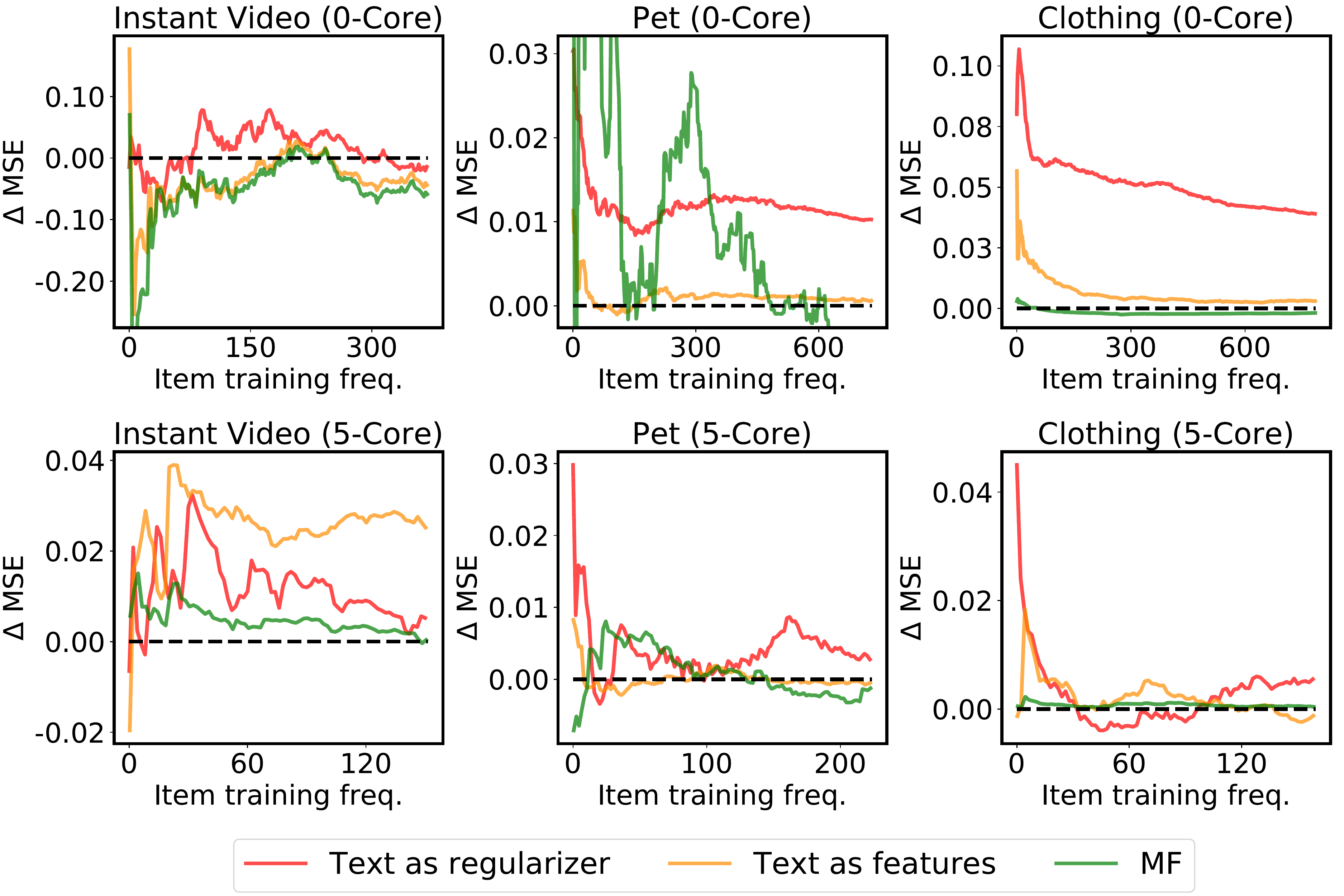}
    \vspace{-0.15cm}
    \caption{Improvement in MSE (higher is better) w.r.t.~the bias only model. Values are smoothed via moving average.}
    \label{freq}
    \vspace{-0.25cm}
\end{figure}

\paragraph{When do reviews help?} In this experiment, we evaluate what part of the item coldness spectrum are the reviews most helpful. We group items based on their training-set frequencies and compare the \textbf{improvement in test-set MSE} (higher is better) of different methods compared to the bias only method. As we can observe (Figure~\ref{freq}), text-based methods tend to differ the most for colder items (left-side of $x$-axis) It is also evident that HFT tends to outperform feature-extraction based methods on 0-core datasets, whereas the opposite generally holds for the 5-core datasets.


\begin{figure}[ht]
    \centering
    \includegraphics[width=8.3cm, height=5.7cm]{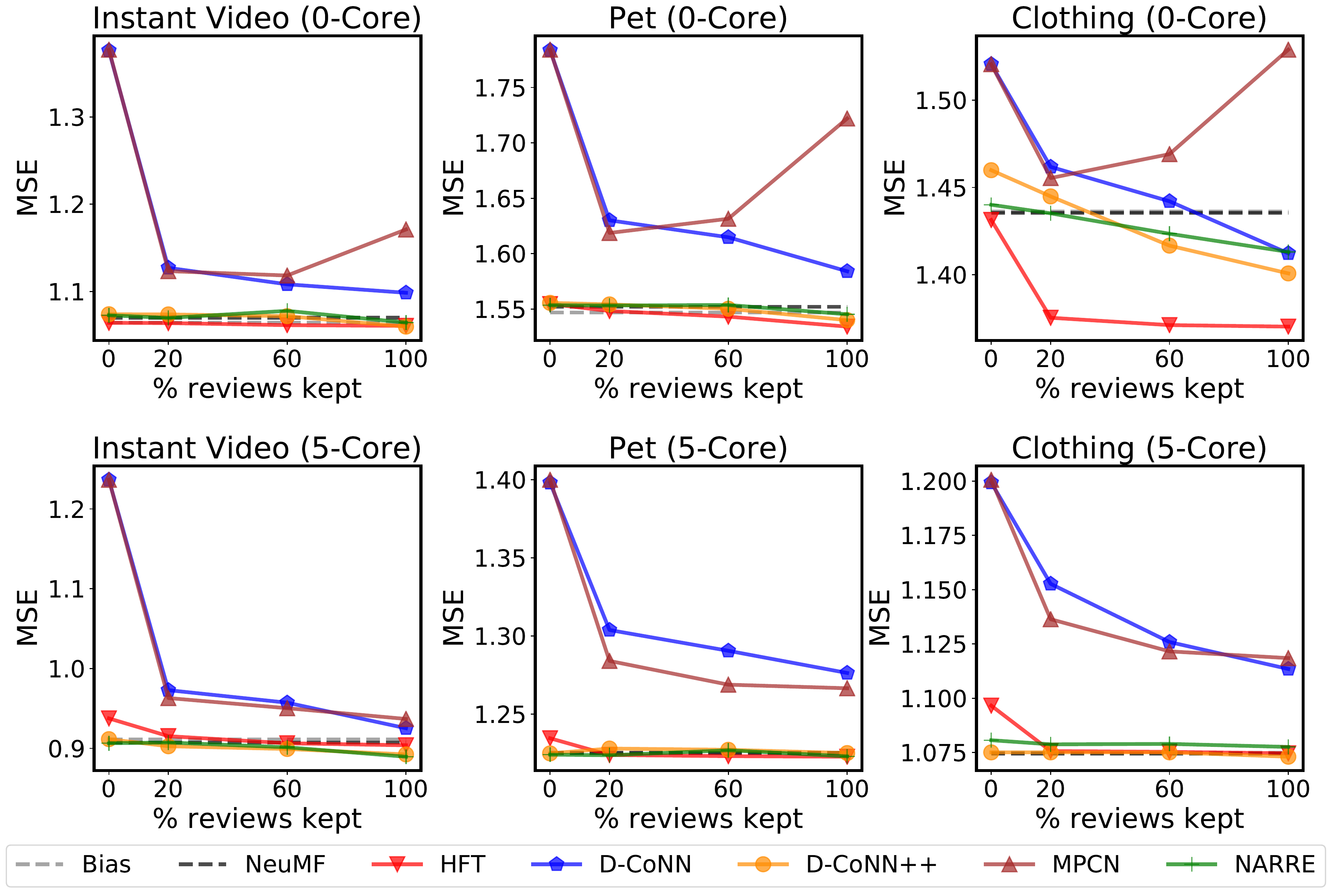}
    \vspace{-0.15cm}
    \caption{Performance comparison when some $\%$ of the reviews are randomly masked.}
    \label{perc_reviews}
    \vspace{-0.35cm}
\end{figure}

\paragraph{How much do reviews help?} To measure the importance of reviews, we propose a simple experiment where we randomly mask $x\%$ of reviews in our dataset to be an empty/null string. On this modified dataset, we train all the methods, varying $x$. In Figure~\ref{perc_reviews}, we observe that methods that rely only on reviews like DeepCoNN, and MPCN degrade vigorously as we randomly remove reviews. On the other hand, methods like DeepCoNN++ and NARRE tend to be relatively unaffected. We conjecture that this behavior arises because of the bias component in DeepCoNN++ and NARRE.


\subsection{Implications}
Through our analysis, we observed interesting insights, anomalies:

\begin{itemize}
    \item There is a relatively small difference in MSE among the bias-only model compared with MF on the 0-core subsets.
    \item Recently published methods like DeepCoNN, NARRE, MPCN, etc.~fail to outperform simpler methods like MF and HFT on both 0, 5-core versions of most review datasets which is in stark contrast of what's expected based on recent literature. 
    \item The major cause of improvement in recent neural methods (eg. NARRE \& DeepCoNN++) is the inclusion of the user and item bias terms rather then their architecture.
    \item The higher capacity, deep-learning-based models surprisingly yield minimal changes in performance when reviews are masked compared to the original setting.
\end{itemize}
In the upcoming section, we suggest possible reasons for these observations, backed with additional experiments.

\vspace{-0.15cm}
\subsection{Discussion}

\paragraph{Reproducibility and Correctness.} Official results mentioned in
most papers considered use 5-core versions of the datasets, which is a dense (and arguably unrealistic) setting of the data, and somewhat against the original motivation for this line of research. We also note that many of the papers considered copy results of competitor methods, despite having changes to data preprocessing strategies. Another problem is incomplete hyper-parameter search. All recent papers show significant gaps among MF-based methods and proposed review-based method, which we find is not supported.

\vspace{-0.07cm}
\paragraph{Overfitting.} We conjecture overfitting to be the main barrier in using highly sophisticated models on unprocessed, existing review datasets. We can see evidence of overfitting in our experiments by observing (1) the \emph{relative} increase in performance of DeepCoNN++, NARRE, and NeuMF compared to bias-only and HFT as we increase the density of our datasets (Figure~\ref{density}); and (2) MF tending to have worse performance than bias-only for low-frequency items (see plots for 0-core datasets in Figure~\ref{freq}).


\vspace{-0.07cm}
\paragraph{Reviews are better used as a regularizer.} In continuation to our previous note about overfitting, we believe reviews are better at regularizing latent factors rather than as more data to extract better features from, especially in cold scenarios. Our belief is supported by the fact that simpler models like HFT perform better on colder items than DeepCoNN(++), NARRE, and MPCN (Figure~\ref{freq}) -- all of which employ reviews to model user/item latent features. We also want to re-iterate that our hypothesis stands only under relatively colder conditions, and more expressive methods like DeepCoNN++ start performing \emph{relatively} better as data density increases (Figure~\ref{density}).

\paragraph{Is MSE at fault?} We could argue that there indeed is an increase in recommendation performance with the newly proposed models, but that our evaluation criterion (MSE) is limited and we should consider more relevant ranking metrics. To assess this, we conduct another experiment: Let $\mathbf{I}_{+}^u$ be the set of items that (test) user $u$ has marked as the maximum rating possible, and $\mathbf{I}_{-}^u$ be the set of items that test user $u$ has marked, but not the maximum rating. We randomly sample one item from $\mathbf{I}_{+}^u$ and five items from $\mathbf{I}_{-}^u$ and rank all six items. We calculate $\mathit{HitRate}@1$ on the ranked list, i.e.,~how many times (on average) was the positive item ranked at the top. 
Results are listed in Table~\ref{hr}. We find in most cases that MSE tracks $\mathit{HR}@1$ (despite some outliers) but exclude results for brevity.

\begin{table}[!ht]
    \begin{footnotesize}
    \centering
    \begin{tabular}{c | c || c || c || c | c} \toprule
        \multicolumn{2}{c||}{\multirow{1}{*}{Dataset}} & \multicolumn{1}{c||}{NeuMF} & \multicolumn{1}{c||}{HFT} & \multicolumn{1}{c|}{D-CoNN++} & \multicolumn{1}{c}{NARRE} \\ \midrule \midrule
        
        \multirow{2}{*}{\shortstack{Instant\\Video}} & 0-core & 1.065 / 40.0 & 1.060 / 40.0 & 1.059 / 40.0 & 1.064 / 33.3 \\ 
        & 5-core & 0.907 / 25.0 & 0.902 / 25.0 & 0.892 / 50.0 & 0.889 / 25.0 \\ \midrule
        
        \multirow{2}{*}{Pet} & 0-core & 1.548 / 13.9 & 1.534 / 16.7 & 1.540 / 13.89 & 1.545 / 16.6 \\
        & 5-core & 1.225 / 27.3 & 1.222 / 36.4 & 1.225 / 27.3 & 1.223 / 27.3 \\ \bottomrule
        
    \end{tabular}
    \end{footnotesize}
    \caption{Ranking metric comparisons: MSE / HR$@1$}
    \label{hr}
    \vspace{-0.7cm}
\end{table}

\paragraph{Conclusions.} Through analyzing 
models that combine ratings and reviews, we conclude that reviews \emph{can} be important,
but the current direction the field is progressing needs to be reconsidered. Inconsistencies in the presentation of results, and impractical/unrealistic data settings can hinder or obscure overall progress. We hope that this work encourages the community to conduct sensible and exhaustive empirical evaluations of their propositions.

\paragraph{Acknowledgements.} This work was partly supported by NSF Award \#1750063.




\bibliography{references}
\bibliographystyle{abbrv}

\end{document}

%% file: all_results.tex
\begin{table*}[!ht]
    \begin{footnotesize}
    \begin{center}
        \begin{tabular}{c | c | c || c c c | c | c c c c c c}
            \toprule
            \multicolumn{3}{c||}{\multirow{3}{*}{\shortstack{Dataset\\\#Reviews / \#Users / \#Items}}} & \multicolumn{3}{c|}{Non-text-based} & Text as regularizer & \multicolumn{6}{c}{Text as features} \\
            
            \multicolumn{3}{c||}{\multirow{2}{*}{}} & \multicolumn{3}{c|}{} & & \multicolumn{6}{c}{} \vspace{-1mm} \\
            
            \multicolumn{3}{c||}{}                                              & Bias        & MF          & NeuMF       & HFT         & D-CoNN      & D-CoNN++    & T-Nets      & T-Nets++ & MPCN    & NARRE \\ \midrule \midrule
            \multirow{2}{*}{Clothing}       & 0-core & 5.7 M / 3.1M / 1.1M      & 1.4362      & 1.4362      & 1.4354      & \bb{1.3703} & 1.4123      & 1.4029      & 1.4730      & 1.4400   & 1.5691   & 1.4131 \\
                                            & 5-core & 0.27M / 39 K / 23 K      & 1.0749      & 1.0749      & 1.0745      & \bb{1.0608} & 1.1135      & 1.0731      & 1.1697      & 1.0793   & 1.1185   & 1.0776 \\ \midrule
            
            \multirow{2}{*}{Toys \& Games}  & 0-core & 2.25M / 1.3M / 0.3M      & 1.3763      & 1.3762      & 1.3776      & \bb{1.3199} & 1.3815      & 1.3216      & 1.4610      & 1.3650   & 1.4817   & 1.3325 \\
                                            & 5-core & 0.16M / 19 K / 12 K      & 0.7926      & 0.7926      & 0.7929      & 0.7890      & 0.8301      & \bb{0.7864} & 0.9278      & 0.7888   & 0.8203   & 0.7913 \\ \midrule
            
            \multirow{2}{*}{Video Games}    & 0-core & 1.32M / 0.8M / 50 K      & 1.5429      & 1.5430      & 1.5432      & \bb{1.5311} & 1.5829      & 1.5320      & 1.6841      & 1.5794   & 1.7448   & 1.5393 \\
                                            & 5-core & 0.23M / 24 K / 10 K      & 1.0962      & 1.0962      & 1.0965      & 1.0914      & 1.1496      & 1.0906      & 1.3721      & 1.1006   & 1.1254   & \bb{1.0882} \\ \midrule
            
            \multirow{2}{*}{Pet}            & 0-core & 1.23M / 0.7M / 0.1M      & 1.5478      & 1.5478      & 1.5482      & \bb{1.5341} & 1.5841      & 1.5400      & 1.5947      & 1.5881   & 1.7308   & 1.5453 \\
                                            & 5-core & 0.15M / 20 K / 8.5K      & 1.2248      & 1.2247      & 1.2252      & \bb{1.2220} & 1.2763      & 1.2250      & 1.3733      & 1.2333   & 1.2665   & 1.2229 \\ \midrule
            
            \multirow{2}{*}{Baby}           & 0-core & 0.91M / 0.5M / 64 K      & 1.4324      & 1.4323      & 1.4328      & \bb{1.4221} & 1.4602      & 1.4242      & 1.4689      & 1.4733   & 1.6209   & 1.4320 \\
                                            & 5-core & 0.16M / 19 K / 07 K      & 1.1282      & 1.1283      & 1.1304      & \bb{1.1212} & 1.1782      & 1.1291      & 1.2617      & 1.1454   & 1.1608   & 1.1260 \\ \midrule
            
            \multirow{2}{*}{Instant Video}  & 0-core & 583K / 0.4M / 24 K       & 1.0643      & 1.0644      & 1.0645      & 1.0605      & 1.0985      & \bb{1.0597} & 1.0884      & 1.1028   & 1.1711   & 1.0643 \\
                                            & 5-core & 37 K / 05 K / 1.6K       & 0.9113      & 0.9088      & 0.9073      & 0.9019      & 0.9252      & 0.8924      & 0.9640      & 0.9168   & 0.9368   & \bb{0.8895} \\ \midrule

            \multirow{2}{*}{BeerAdvocate}   & 0-core & 1.58M / 33 K / 66 K      & 0.3709      & 0.3688      & 0.3667      & \bb{0.3605} & 0.3808      & 0.3705      & 0.4333      & 0.3805   & 0.3715   & 0.3648 \\
                                            & 5-core & 1.47M / 15 K / 22 K      & 0.3561      & 0.3538      & 0.3513      & \bb{0.3477} & 0.3684      & 0.3562      & 0.4173      & 0.3617   & 0.3585   & 0.3503 \\ \bottomrule 
        \end{tabular}
    \end{center}
    \end{footnotesize}
    \bigskip
    \caption{Data statistics (left), and MSE values (right) of various methods. Bold values represent the best method in that row.}
    \label{all_results}
    \vspace{-6mm} 
\end{table*}

%% file: sample-authordraft.bbl
\begin{thebibliography}{10}

\bibitem{lda}
D.~M. Blei, A.~Y. Ng, and M.~I. Jordan.
\newblock Latent dirichlet allocation.
\newblock {\em Journal of Machine Learning Research}, 2003.

\bibitem{transnet}
R.~Catherine and W.~Cohen.
\newblock Transnets: Learning to transform for recommendation.
\newblock In {\em ACM RecSys}, 2017.

\bibitem{narre}
C.~Chen, M.~Zhang, Y.~Liu, and S.~Ma.
\newblock Neural attentional rating regression with review-level explanations.
\newblock In {\em WWW}, 2018.

\bibitem{making_progress}
M.~F. Dacrema, P.~Cremonesi, and D.~Jannach.
\newblock Are we really making much progress?
\newblock In {\em ACM RecSys}, 2019.

\bibitem{dataset_paper}
R.~He and J.~McAuley.
\newblock Ups and downs: Modeling the visual evolution of fashion trends with
  one-class collaborative filtering.
\newblock In {\em WWW}, 2016.

\bibitem{neural_mf}
X.~He, L.~Liao, H.~Zhang, L.~Nie, X.~Hu, and T.-S. Chua.
\newblock Neural collaborative filtering.
\newblock In {\em WWW}, 2017.

\bibitem{textcnn}
Y.~Kim.
\newblock Conv. neural networks for sentence classification.
\newblock In {\em EMNLP}, 2014.

\bibitem{mf}
Y.~Koren, R.~Bell, and C.~Volinsky.
\newblock Matrix factorization techniques for recommender systems.
\newblock {\em Computer}, 42(8), 2009.

\bibitem{hft}
J.~McAuley and J.~Leskovec.
\newblock Hidden factors and hidden topics: Understanding rating dimensions
  with review text.
\newblock In {\em ACM RecSys}, 2013.

\bibitem{beer_dataset}
J.~McAuley, J.~Leskovec, and D.~Jurafsky.
\newblock Learning attitudes and attributes from multi-aspect reviews.
\newblock In {\em IEEE ICDM}, 2012.

\bibitem{mpcn}
Y.~Tay, A.~T. Luu, and S.~C. Hui.
\newblock Multi-pointer co-attention networks for recommendation.
\newblock In {\em ACM SIGKDD}, 2018.

\bibitem{mengting_goodreads}
M.~Wan and J.~McAuley.
\newblock Item recommendation on monotonic behavior chains.
\newblock In {\em ACM RecSys}, 2018.

\bibitem{deepconn}
L.~Zheng, V.~Noroozi, and P.~S. Yu.
\newblock Joint deep modeling of users and items using reviews for
  recommendation.
\newblock In {\em ACM WSDM}, 2017.

\end{thebibliography}
